\documentclass[11pt]{article}
\usepackage[utf8]{inputenc}
\usepackage[T1]{fontenc}
\usepackage{lmodern}
\usepackage[margin=1in]{geometry}
\usepackage{amsmath,amssymb}
\usepackage{graphicx}
\usepackage{booktabs}
\usepackage{hyperref}
\usepackage{xcolor}
\usepackage{natbib}
\usepackage{enumitem}
\usepackage{placeins}
\usepackage{url}
\usepackage{microtype}
\hypersetup{colorlinks=true, linkcolor=blue!70!black, citecolor=blue!70!black, urlcolor=blue!70!black}
\newcommand{\lstar}{\lambda^\star}
\newcommand{\Fz}{F_0}

\title{Group Commit Self-Clocks:\\ Why Tuning Is Unnecessary Above a Device-Set Load Threshold}
\author{Madhulatha Mandarapu\thanks{madhulatha@samyama.ai} \and Sandeep Kunkunuru\thanks{sandeep@samyama.ai}}
\date{VaidhyaMegha Private Limited, India\\[2pt]\url{https://samyama.ai/}\\[8pt]June 2026}

\begin{document}
\maketitle

\begin{abstract}
Group commit amortizes the fixed cost of a durable log flush across many committing transactions; the
release rule -- a timer, a batch size, or an adaptive policy -- is a classic tuning knob. The textbook
theory is open-loop: for Poisson arrivals the optimal timer is the EOQ square-root rule, and the
wait-or-flush decision is ski-rental 2-competitive. We ask when that tuning is worth its machinery, and
show that in closed-loop OLTP it usually is not. Real
commit arrivals are closed-loop: a client issues its next transaction only after its last commits, so the
arrival rate is induced by the policy's own latency. Modeling this as a closed queueing network, the
parameter-free greedy-pipelined policy (flush the instant the device is free) self-clocks to a computable
fixed point and is within about $0.1\%$ of the best oracle-tuned timer at every load. The square-root rule
prescribes waiting $T^\star=\sqrt{2\Fz/\lambda}$, but $T^\star<\Fz$ exactly when $\lambda>\lstar=2/\Fz$;
above this device-set load threshold the timer collapses onto greedy and tuning is vacuous. The clean
theory only bites below $\lstar$ and in the open-loop world, where a parameter-free ski policy still beats
a fixed tuned timer under rate shifts. We instantiate $\lstar$ with measured \texttt{fsync} distributions
on two AWS storage classes (EBS gp3 versus instance NVMe, a $25\times$ range), and confirm on PostgreSQL
that \texttt{commit\_delay}=0 is competitive with any tuned value. The contribution is a characterization
that explains deployed practice; we add no new logger.
\end{abstract}

\section{Introduction}
A durable transaction commit must force the write-ahead log to stable storage with a flush
(\texttt{fsync}, an NVMe barrier, or a replicated quorum acknowledgment). Because a flush has a large
fixed cost, \emph{group commit}~\citep{dewitt1984mmdb,gawlick1985imsfastpath} batches many committing
transactions into one flush. The \emph{release rule} -- flush after a timer $T$, after $K$ transactions
queue, or by some adaptive rule -- trades the first transaction's \emph{acquisition delay} against the
\emph{amortization} of the flush over the batch. Every serious engine exposes this as a tuning knob
(PostgreSQL \texttt{commit\_delay}/\texttt{commit\_siblings}, MySQL binlog group commit), and a long line
of work optimizes it~\citep{johnson2010aether,deb1973batch}.

The textbook analysis is \emph{open-loop}: model commits as exogenous Poisson arrivals at rate $\lambda$
with fixed flush cost $\Fz$. The per-transaction overhead of a periodic timer is $\Fz/(\lambda T)+T/2$,
minimized at the EOQ square-root rule $T^\star=\sqrt{2\Fz/\lambda}$; and the wait-or-flush decision is the
rent-or-buy / ski-rental problem~\citep{karlin1994skirental}, equivalently the dynamic-TCP-acknowledgment
problem~\citep{dooly2001tcp}, which is $2$-competitive deterministic. We ask a question upstream of all the
tuning: \emph{when is the tuning worth its machinery?}

Our answer is that, in the regime real OLTP systems run in, it largely is not -- and we make precise why.
Real commit arrivals are \emph{closed-loop}: with a bounded connection pool, a client issues its next
transaction only after its previous one commits, so $\lambda$ is a function of commit latency
(``latency throttles the very arrivals it batches''). The open-loop theory discards exactly this. We
contribute (\S\ref{sec:theory}--\ref{sec:exp}):
\begin{enumerate}[leftmargin=1.4em]\itemsep1pt
  \item \textbf{Self-clocking fixed point.} Modeled as a closed queueing network, the parameter-free
        greedy-pipelined policy converges to a computable batch/throughput fixed point and saturates the
        device.
  \item \textbf{Parameter-free optimality.} Greedy is within $\sim\!0.1\%$ of the best oracle-tuned timer
        at every closed-loop load -- no $\lambda$, no timer.
  \item \textbf{The load threshold $\lstar=2/\Fz$.} The square-root rule's prescription becomes
        \emph{vacuous} above $\lstar$: the timer collapses onto greedy. Tuning can only help below it.
  \item \textbf{Real instantiation.} Measured \texttt{fsync} on two AWS storage classes sets $\lstar$
        (2.2k/s vs 55k/s); a PostgreSQL \texttt{commit\_delay} sweep shows no reliable tuning signal.
\end{enumerate}
We claim no new logger. Flush pipelining~\citep{johnson2010aether} is the mechanism that makes
self-clocking possible; our contribution is the analysis that explains the deployed ``set
\texttt{commit\_delay}$\approx 0$'' practice (\S\ref{sec:related}, \S\ref{sec:limits}).

\section{Model and theory}\label{sec:theory}
A flush of a batch of $K$ records costs $F(K)=\Fz+K\delta$ (fixed barrier $\Fz$ plus per-record $\delta$).
The controllable per-transaction cost is $\Fz/\bar K + \overline{\text{latency}}$. \textbf{Open loop:}
exogenous Poisson($\lambda$) gives $T^\star=\sqrt{2\Fz/\lambda}$ and the ski-rental $2$-competitive bound.
\textbf{Closed loop:} $N$ clients, each thinking for $Z$ then blocking until its commit's flush completes.
Greedy-pipelined flushes the instant the device is free, batching whatever queued. By Little's law
$X=N/(Z+L)$ and, with the device back-to-back at load, $X=K/F$ and $L\approx\rho F$; eliminating gives the
fixed point
\begin{equation}\label{eq:fp}
  K^\star=\frac{N\,F(K^\star)}{Z+\rho\,F(K^\star)},\qquad X\to 1/\delta\ \text{(write-bandwidth bound)} .
\end{equation}
The key observation is the \textbf{threshold}: $T^\star=\sqrt{2\Fz/\lambda}<\Fz \iff \lambda>\lstar:=2/\Fz$.
Above $\lstar$ the prescribed wait $T^\star$ is shorter than a single flush, so a timer fires while the
device is still busy and the batch flushes at device-free time regardless of $T$ -- i.e.\ the timer
\emph{is} greedy. Tuning the timer is therefore vacuous above $\lstar$; it can only matter at low load
($\lambda<\lstar$) or in the open-loop world.

\section{Experiments}\label{sec:exp}
All hypotheses and decision rules were pre-registered before the real-data runs; code, seeds, and the
pre-registration (including two hypotheses the data reframed) are public.\footnote{\url{https://github.com/samyama-ai/group-commit-policy}}
We use a discrete-event simulator with a flush device, an offline dynamic-acknowledgment DP (verified
against brute force), and the policies above.

\begin{figure}[t]\centering
  \begin{minipage}{0.62\linewidth}\includegraphics[width=\linewidth]{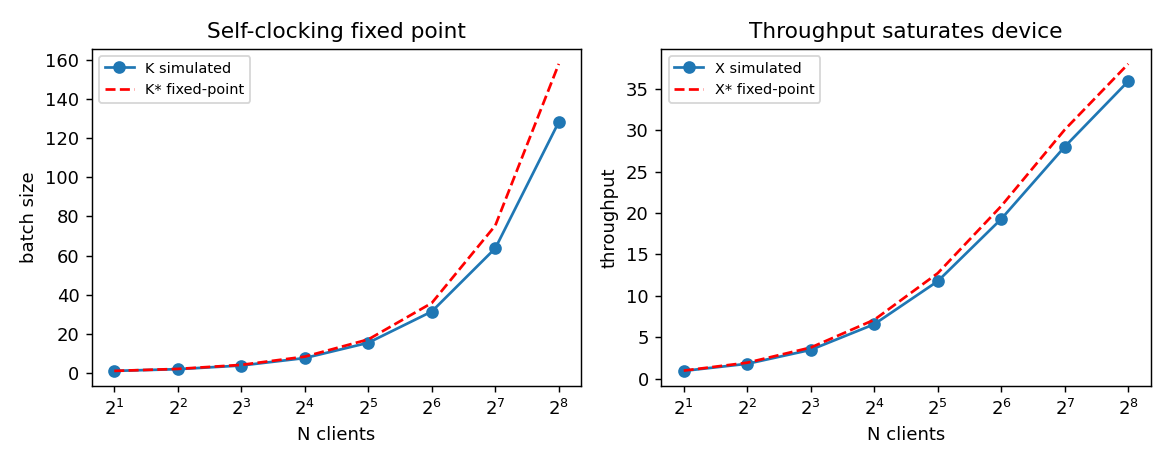}
    \caption{Self-clocking fixed point: greedy batch and throughput vs.\ the closed-network
      prediction.}\label{fig:fp}\end{minipage}\hfill
  \begin{minipage}{0.36\linewidth}\includegraphics[width=\linewidth]{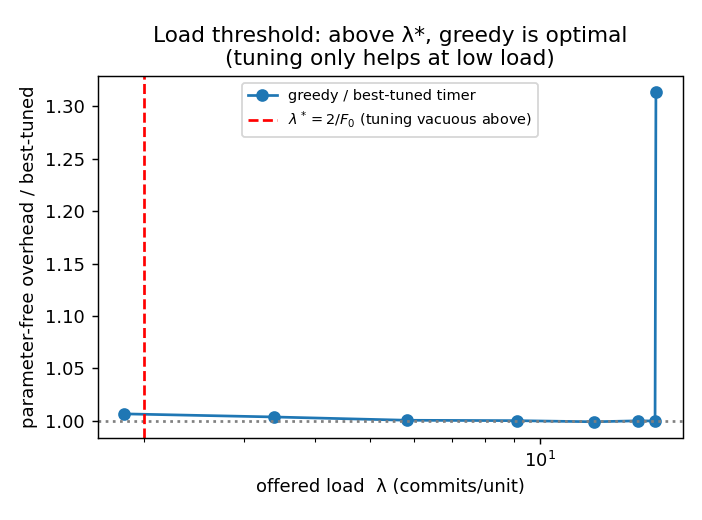}
    \caption{Above $\lstar=2/\Fz$ tuning is vacuous.}\label{fig:thr}\end{minipage}
\end{figure}

\paragraph{Self-clocking fixed point (Fig.~\ref{fig:fp}).} Greedy's batch size and throughput match
\eqref{eq:fp} across $N\in\{2,\dots,256\}$ (batch error $10\%$, throughput error $7\%$); the device
saturates.

\paragraph{Parameter-free optimality and the threshold (Fig.~\ref{fig:thr}).} Across every closed-loop
load, greedy is within a median $\mathbf{0.1\%}$ of the best oracle-tuned timer. The open-loop
square-root-rule timer, fed the measured closed-loop $\lambda$, gives \emph{exactly} greedy's overhead
(ratio $1.000$) for all $\lambda>\lstar$, and diverges only below $\lstar$ -- confirming the threshold.

\begin{figure}[t]\centering
  \begin{minipage}{0.34\linewidth}\includegraphics[width=\linewidth]{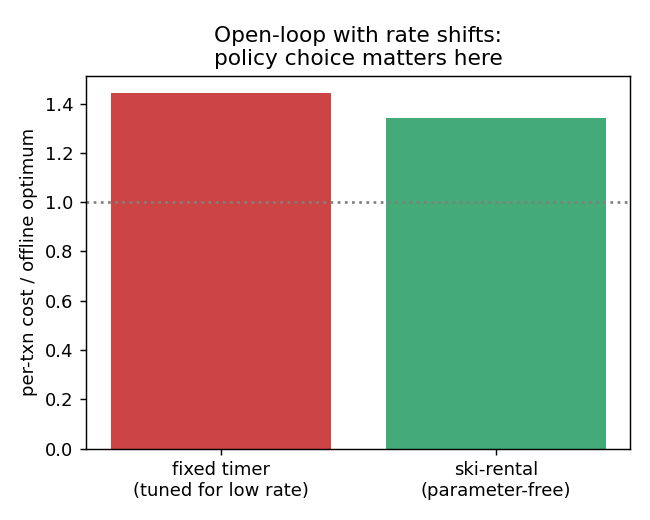}
    \caption{Open-loop rate shifts: policy matters.}\label{fig:con}\end{minipage}\hfill
  \begin{minipage}{0.64\linewidth}\includegraphics[width=\linewidth]{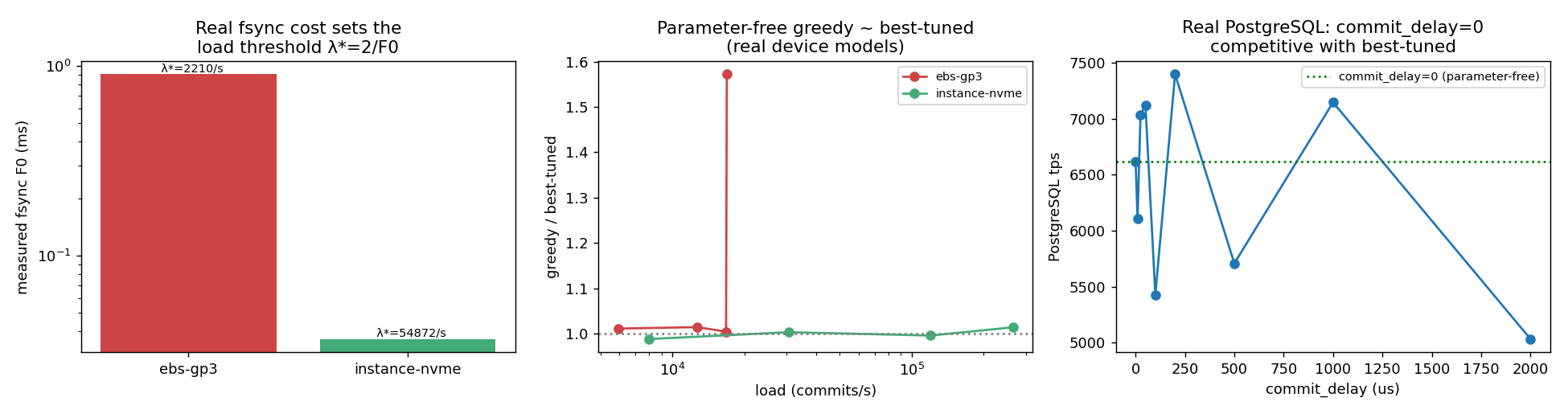}
    \caption{Real \texttt{fsync} sets $\lstar$; greedy $\approx$ best-tuned; PostgreSQL
      \texttt{commit\_delay}=0 competitive.}\label{fig:dev}\end{minipage}
\end{figure}

\paragraph{Where policy still matters (Fig.~\ref{fig:con}).} In the open-loop abstraction with rate
shifts, a fixed timer tuned for the low rate is hit by bursts; a parameter-free ski-rental policy adapts.
Ski lands at $1.34\times$ the offline optimum versus the fixed timer's $1.44\times$ -- modest but real,
and it locates the regime where the clean theory bites.

\paragraph{Real devices and PostgreSQL (Fig.~\ref{fig:dev}).} We measure the \texttt{fsync} latency
distribution on two AWS storage classes. EBS gp3 has $\Fz=0.90$\,ms (p99 $2.5$\,ms), so
$\lstar\approx 2200$/s -- any non-trivial OLTP load is above it, and group commit matters. Instance-store
NVMe has $\Fz=0.036$\,ms ($25\times$ faster), so $\lstar\approx 55000$/s -- group commit barely matters
until extreme load, quantifying the ``is group commit needed on fast storage?'' question. Plugging each
measured distribution into the simulator, greedy is within $1.3\%$ (EBS) / $0.1\%$ (NVMe) of best-tuned.
Finally, a PostgreSQL~\citep{postgresql} \texttt{commit\_delay} sweep (pgbench, WAL on EBS): at $32$ clients ($\sim$6.6k
tps, above $\lstar$) \texttt{commit\_delay}=0 is within $\sim$10\% of the best tuned value with no reliable
or monotonic signal; at $2$ clients (below $\lstar$) the benefit is slightly larger ($\sim$17\%), the
direction the threshold predicts. The practical upshot matches deployed folklore: leave
\texttt{commit\_delay}$\approx 0$ and let the log self-clock.


\FloatBarrier   
\section{Related work}\label{sec:related}
Group commit originates with \citet{dewitt1984mmdb} and \citet{gawlick1985imsfastpath}. \emph{Aether}
\citep{johnson2010aether} introduced flush pipelining -- agents commit asynchronously without blocking on
the flush -- which is the \emph{mechanism} that lets the log self-clock; our contribution is the
\emph{analysis} of the resulting closed-loop policy, not the mechanism. \citet{deb1973batch} give the
open-loop bulk-service optimum under known statistics; ski-rental~\citep{karlin1994skirental} and dynamic
acknowledgment~\citep{dooly2001tcp} give the rent-or-buy competitive theory we use for the open-loop
regime. Cloud log services~\citep{verbitski2017aurora} change $\Fz$ to a network-tail distribution,
shifting $\lstar$ but not the structure. The broader point that commit I/O is a dominant OLTP
cost~\citep{harizopoulos2008oltp} is what makes the regime question matter.

\section{Limitations and honest scope}\label{sec:limits}
The headline is a \emph{debunking}: in closed-loop OLTP, adaptive group-commit tuning is largely
unnecessary, and we explain precisely why. That matches deployed practice; it is a characterization, not a
new record, and the mechanism is Aether's. Two pre-registered hypotheses (a strong open-loop misprediction
and burst-fragility of timers) were \emph{not} supported as stated; the data reframed them into the
threshold result above, disclosed in our pre-registration. The closed-loop competitive bound is
conjectured, not proven; we report the empirical optimality gap. pgbench has non-commit bottlenecks and
spot/EBS latency is variable, so the PostgreSQL arm is noisy -- we report that rather than cherry-pick.
Every number is regenerated by one command over public code and measured device costs.

\small
\bibliographystyle{plainnat}
\bibliography{paper12_group_commit}
\end{document}